# LBL System for Underwater Acoustic Positioning: Concept and Equations

Pablo Otero, Álvaro Hernández-Romero, and Miguel-Ángel Luque-Nieto

*Abstract*— GNSS (Global Navigation Satellite System) positioning is not available underwater due to the very short range of electromagnetic waves in the sea water medium. In this article a LBL (Long Base Line) acoustic repeater system of the GNSS positioning is presented. The system is hyperbolic, i.e., based on time differences and it doesn't need very accurate atomic clocks to synchronize repeaters.

## I. Introduction

OCEANS and seas are threatened due to always growing human activity. Pollution is the most dangerous threat to sea life. Less polluting human activities and more efficient sea monitoring are the target of many oceanic engineering researchers. On the other hand, oceans exploitation is needed for human food, minerals extraction and transportation. AUVs (Autonomous Underwater Vehicles) play a paramount role in the sustainable exploitation of the seas, that will protect water quality and sea life, while fairly using the resources they keep inside.

AUVs will only be possible if an underwater positioning system is available. GNSS (Global Navigation Satellite System) cannot provide that service due to the high absorption of electromagnetic waves in the salted water media. Some research effort has been devoted to underwater positioning. Existing systems are commonly classified according to their range: USBL (Ultra Short Base Line), SBL (Short Base Line) and LBL (Long Base Line). This article proposes a LBL system based on classical hyperbolic systems that uses GNSS service available on the sea surface to broadcast signals to the sea bottom that will allow a receiver to calculate its position. The system does not require accurate atomic clocks. There is also no limitation on the number of receivers that can be serviced.

The main error sources are the intrinsic GNSS error (which is also present in the regular GNSS service), and the acoustic wave refraction due to the variable vertical profile of the sound propagation velocity in seawater, as well as the variable velocity itself.

Other errors are of an order of magnitude less than that due to refraction. This article deals with the system concept and equations. Refraction error and its compensation will be considered later.

In this article, Section II contains the description of the proposed system, Section III presents the equations that allow an underwater receiver to compute its position, and Section IV contains the results of a test to verify the proper system operation, as well as a discussion on the system limitations. Finally, the article contents will be summarized in Section V.

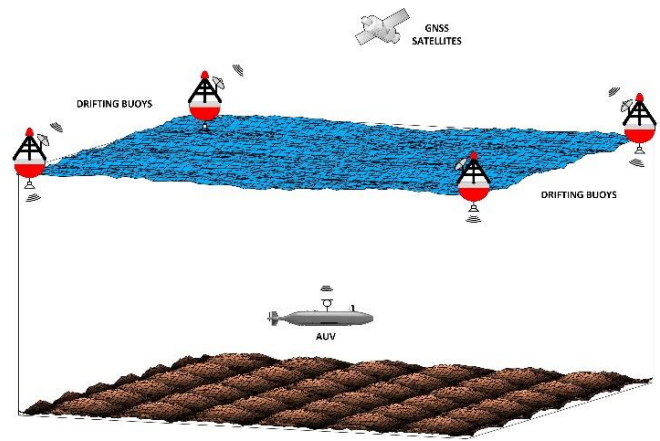

**Fig. 1.** System sketch.

## II. System description

The proposed system consists of at least four surface buoys that transmit messages containing their GNSS-position and the GNSS-time at which the buoy position was computed. Fig.1 shows a schematic view of the system. The buoys positions define a quadrilateral on the sea surface. The preferred quadrilateral is a square, but the system is still possible if the buoys are not online.

Surface buoys can be drifting buoys, as it will be soon understood, in Section III. The reason lies in the fact that each buoy transmits its position at a particular instant, which is all the analytical solution needs to calculate its position.

Each buoy transmits sequentially, through an underwater acoustic transmitter, a message containing its GNSS position and the time instant at which that position was calculated. All buoys use the GNSS time, so that we can consider that they are synchronized. All receivers that are in the acoustic communication range of four buoys can calculate their position. This article deals with the multilateration equations solution; however, it is mandatory to demonstrate the feasibility of the time schedule of buoys transmissions. The length of the message is estimated to be around 80 bytes. This number comes from the length of the complete frame of the standard NMEA [1] and could be shortened if needed. Assuming a maximum limit of one second for the transmission time of a message, the minimum bit rate for the transmitter would be 640 bps, which is achievable with commercial underwater acoustic modems. Concerning propagation time, one second roughly means 1.5 km (the propagation speed depends on the salinity and temperature of the water and on depth), which is a long range in the world of underwater communications. A guard time of one second can be considered so far, which means a duty cycle of less than 10 s for the four buoys and a delay of around 8 s between the transmission of buoy #1 and the transmission of buoy #4. Fig. 2 shows the frame structure.

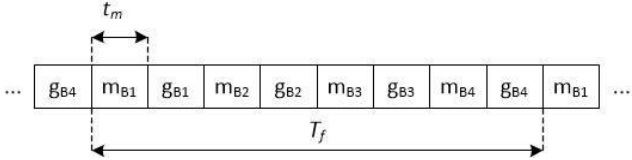

**Fig. 2.** Frame structure.

In Fig. 2, $m_{Bi}$ stands for the message transmitted by buoy $i$, and $g_{Bi}$ is the guard time before buoy $i+1$ starts transmitting; $t_m$ is the duration of the buoy message and $T_f$ is the length of a single frame, which is the period of the transmission of the four buoys. On the other hand, an AUV is unlikely to move at more than 10 knots, which is around 5 m/s. In 8 s, it would have moved 40 m. This figure can be roughly considered the upper bound of the intrinsic limitation of the system in terms of position accuracy. Nevertheless, in case a lower upper limit is required, the use of a higher data rate in the modem or the reduction of the guard time results in a shorter frame and, consequently, higher position accuracy.

### III. ANALYTIC SOLUTION OF THE MULTILATERATION EQUATIONS

The analytical solution of the so-called multilateration problem is presented below. So far, a constant sound propagation velocity $c$ is assumed. As mentioned, the four buoys use the GNSS clock. Let $t_i$ be the instant of transmission of buoy $i = 1, ..., 4$, and $t'_i$ the instant of the reception of the corresponding message by the underwater receiver. The message transmitted by buoy $i$ contains $t_i$ and the GNSS-position of the buoy, $\mathbf{R}_i$.

Attention must be paid to the fact that the clock of $t_i$ is the GNSS-time at buoy i and the clock of $t'_i$ is the underwater receiver clock, which does not need to be synchronized with the GNSS time. That means that $t_i$ and $t'_i$ are measured by different clocks. Let $\Delta t$ be the offset between the two clocks. The so-called pseudorange between buoy $i$ and the receiver is calculated as

$$P_i = c(t'_i - t_i + \Delta t) \tag{1}$$

The difference of pseudoranges between buoys $i$ and $j$ and the receiver is

$$d_{ji} = P_j - P_i = c[(t'_j - t_j) - (t'_i - t_i)], \tag{2}$$

because the offset in the pseudoranges cancels out. That is, the receiver knows $\mathbf{R}_i$ and $d_{ji}$, which are the data that are needed to solve the multilateration problem. We particularly liked the analytical solution of Kleusberg [2]. Final equations of Kleusberg's solution are provided below for the sake of completeness. The position of one the buoys is used as a reference and is denoted by $\mathbf{R}_0$. Let us use $\mathbf{R}^{1,2}$ for the Kleusberg's analytical solution for the receiver position. There are two solutions, identified with the superscript, in the same vertical axis; the sought solution is the underwater one. The position of the receiver is referred to the reference buoy as

$$\mathbf{R}^{1,2} = \mathbf{R}_0 + \mathbf{e}^{1,2} \, s_0^{1,2}, \tag{3}$$

where the distance $s_0^{1,2}$ is calculated as

$$s_0^{1,2} = \frac{1}{2} \frac{b_{0i}^2 - d_{0i}^2}{d_{0i} + b_{0i}(\mathbf{e}^{1,2} \cdot \mathbf{e}_{0i})} . \tag{4}$$

In (4), $d_{0i}$ is the pseudorange difference given by (2), $\mathbf{e}_{0i} b_{0i}$ is the position vector of buoy $i$ from buoy 0, and $\mathbf{e}^{1,2}$ is an unit vector that is calculated as

$$\mathbf{e}^{1,2} = (\mathbf{G} \cdot \mathbf{G})^{-1}\{\mathbf{G} \times \mathbf{H} \pm \mathbf{G}[(\mathbf{G} \cdot \mathbf{G}) - (\mathbf{H} \cdot \mathbf{H})]^{1/2}\}. \tag{5}$$

Vectors $\mathbf{G}$ and $\mathbf{H}$ are calculated as

$$\mathbf{G} = \mathbf{F}_1 \times \mathbf{F}_2, \tag{6}$$

$$\mathbf{H} = U_2 \mathbf{F}_1 - U_1 \mathbf{F}_2, \tag{7}$$

where

$$\mathbf{F}_k = \frac{b_{0k}}{b_{0k}^2 - d_{0k}^2} \mathbf{e}_{0k} - \frac{b_{0,k+1}}{b_{0,k+1}^2 - d_{0,k+1}^2} \mathbf{e}_{0,k+1}, k = 1,2 \tag{8}$$

and

$$U_k = \frac{d_{0,k+1}}{b_{0,k+1}^2 - d_{0,k+1}^2} - \frac{d_{0k}}{b_{0k}^2 - d_{0k}^2}, \quad k = 1,2 \tag{9}$$

From the conceptual point of view, the solution to the multilateration problem consists of finding the intersection of at least three hyperboloids. The intersection of two hyperboloids, if it exists, is a line in space. The lines resulting from intersecting the hyperboloids two by two in a real case do not have to intersect and then the Kleusberg method, which has the elegance of solving the problem analytically, does not provide the point where the receiver is located. The Kleusberg method can be used in the laboratory to test other algorithms, which is what has been done in this work. Numerical multilateration algorithms can be used to calculate the position of actual receivers.

## IV. Results

The analytical solution has been tested using the reverse calculation. The test consists in placing four buoys on the sea surface and a receiver in determined underwater positions. That is, the GNSS positions of the four buoys are known. Pseudoranges are calculated and, eventually, the eight values of $t'_i$ and $t_i$. A stationary receiver has been considered. The four messages are formed and given to the receiver, that computes its position using the analytical solution presented in Section III and it is found to be exactly correct.

## V. Conclusion

In this letter a GNSS repeater system has been presented to provide underwater positioning using four acoustic transmitters mounted on surface buoys. The main contribution of the chart is that the underwater receiver clock does not need to be synchronized with the buoy clocks or be a very accurate atomic clock. The proof of concept has been carried out using the Kleusberg analytical method to solve the multilateration problem, and the feasibility of the time schedule has also been demonstrated.

As mentioned in Section II, drifting of the buoys is not a limitation because the buoy transmits its position at a particular instant, which is all the receiver algorithm needs to calculate the difference of pseudoranges.

Regarding the system limitations, the following can be mentioned. The service range is limited by the communication range of the underwater acoustic communication system. The position accuracy is limited by the transmission bit rate of the above-mentioned system, as explained in Section II. There is no limitation on the number of receivers that the system can serve, as long as they are in the range of four buoys.

## Acknowledgment

The authors would like to thank Prof. Dr. María Ángeles Gómez-Molleda, University of Málaga, who helped them with the multilateration equations.

**Authors' affiliation:** P. Otero (pablo.otero@uma.es), Á. Hernández-Romero (alvaro.h.r@uma.es), and M.-Á. Luque-Nieto (luquen@uma.es) are with the Institute of Oceanic Engineering Research, University of Málaga, Málaga, Spain. *(Corresponding author: M.-Á. Luque-Nieto).*

**Funding:** This research has been funded by a grant from the Spanish Ministry of Science and Innovation through the project "NAUTILUS: Swarms of underwater autonomous vehicles guided by artificial intelligence: its time has come" (PID2020-112502RB / AEI / 10.13039/501100011033).